\documentclass[a4paper,aps,prl,twocolumn,floatfix,showpacs]{revtex4}
\usepackage{amsfonts}
\usepackage{amssymb}
\usepackage{graphicx}
\usepackage{texdraw}
\usepackage{color}
\usepackage{subfigure}

\begin{document}

\title{Controlled switching between paramagnetic and diamagnetic Meissner effect in Pb/Co nanocomposites} \author{Y. T.  Xing, H. Micklitz, E. Baggio-Saitovitch}
\affiliation{Centro Brasileiro de Pesquisas F\'isicas, Rio de Janeiro 22290-180,  Brazil}
\author{T. G. Rappoport} \affiliation{Instituto de  F\'{\i}sica, Universidade Federal do Rio de Janeiro, Cx. P. 68528, 21941-972 , Rio de Janeiro, Brazil}

\date{\today}

\begin{abstract}

A hybrid system which consists of a superconducting (SC) Pb film (100 nm thickness) containing $\sim$1 vol\% single domain ferromagnetic (FM) Co particles of mean-size $\sim$4.5 nm reveal unusual magnetic properties: (i) a controlled switching between the usual diamagnetic and the unusual paramagnetic Meissner effect in field cooling as well as in zero-field cooling experiments (ii) amplification of the positive magnetization when the sample enters the SC state below T$_c$. These experimental findings can be explained by the formation of spontaneous vortices and the possible alignment of these vortices due to the foregoing alignment of the Co particle FM moments by an external magnetic field. 
\end{abstract}


\maketitle


The paramagnetic Meissner effect (PME) has been extensively studied in both conventional and high T$_c$ superconductors \cite{AKGeim98,WBraunish93,DJThompson95,AEKoshelev95,FTGias04,MATorre06,CMonton07}. There are many explanations for the PME such as randomly oriented  $\pi$-junctions in high T$_c$ superconductors\cite{DDominguez94,JMagnusson95}, flux compression \cite{GSOkram97, VVMoshchalkov97}, surface effect\cite{CHeinzel93,SYuan04} and special microstructure \cite{SRiedling94,EKhalil97}. A review article for the PME can be found in the literature\cite{MSLi03}. Chu et al.\cite{SYChu06} reported that if there is a low T$_c$ phase surrounded by a high T$_c$ phase in the sample, the so-called extrinsic PME (EPME) can be observed in FC measurements. Very recently, R. Miller {\it et al.} observed a PME in 18R-SnSe$_2$\{CoCp$_2$\}$_{0.1}$ in ZFC measurements but the origin of it is not clear\cite{RMiller08}.

In this letter we report on a new phenomena: controlled switching between PME and DME with the same external magnetic field in superconductor(SC)/ferromagnet(FM) nanocomposites. This switching can be performed in both field-cooled (FC) and zero-field-cooled (ZFC) measurements. The mechanism of the observed PME in SC/FM nanocomposites is completely different from that observed in all the previous systems: it is exclusively due to the spontaneous vortices induced by the FM nanoparticles embedded in the SC matrix. The different contributions of the external field and the spontaneous vortices to the magnetization of the sample make it possible to manipulate PME and DME by changing the orientation of the nanoparticles' magnetic moments inside the SC.

The sample is a hybrid system consisting of a 100 nm lead (Pb) film with 1\% volume of homogeneously distributed Co particles, following a well established preparation method. These particles, with 4.5 nm diameter, were produced by the so-called inert-gas aggregation method and co-deposited with Pb atoms onto a cold (40 K) sapphire (or quartz) substrate. This method has some advantages: (i) the size of the Co nanoparticles is tunable and its distribution is quite narrow, (ii) the orientation of the magnetic moments of the Co nanoparticles in as-prepared samples is completely random. The Co volume fraction in the Pb matrix  was controlled in-situ by three quartz balances and later checked ex-situ by EDX. After deposition, the samples were annealed at 300 K in order to decrease the lattice defect density. A more detailed description of the experimental set-up and operating procedures can be found in literature~\cite{SRubin98,YTXing08}. After the in-situ transport measurements the sample was taken out of the preparation chamber and immediately put in a quantum-design MPMS-xl SQUID for the magnetic measurements to avoid the  oxidation of Pb. The external magnetic field in all measurements was parallel to the surface of the film. Both FC and ZFC measurements were performed in a warming up process.

\begin{figure}[h]
\includegraphics[width=0.8\columnwidth]{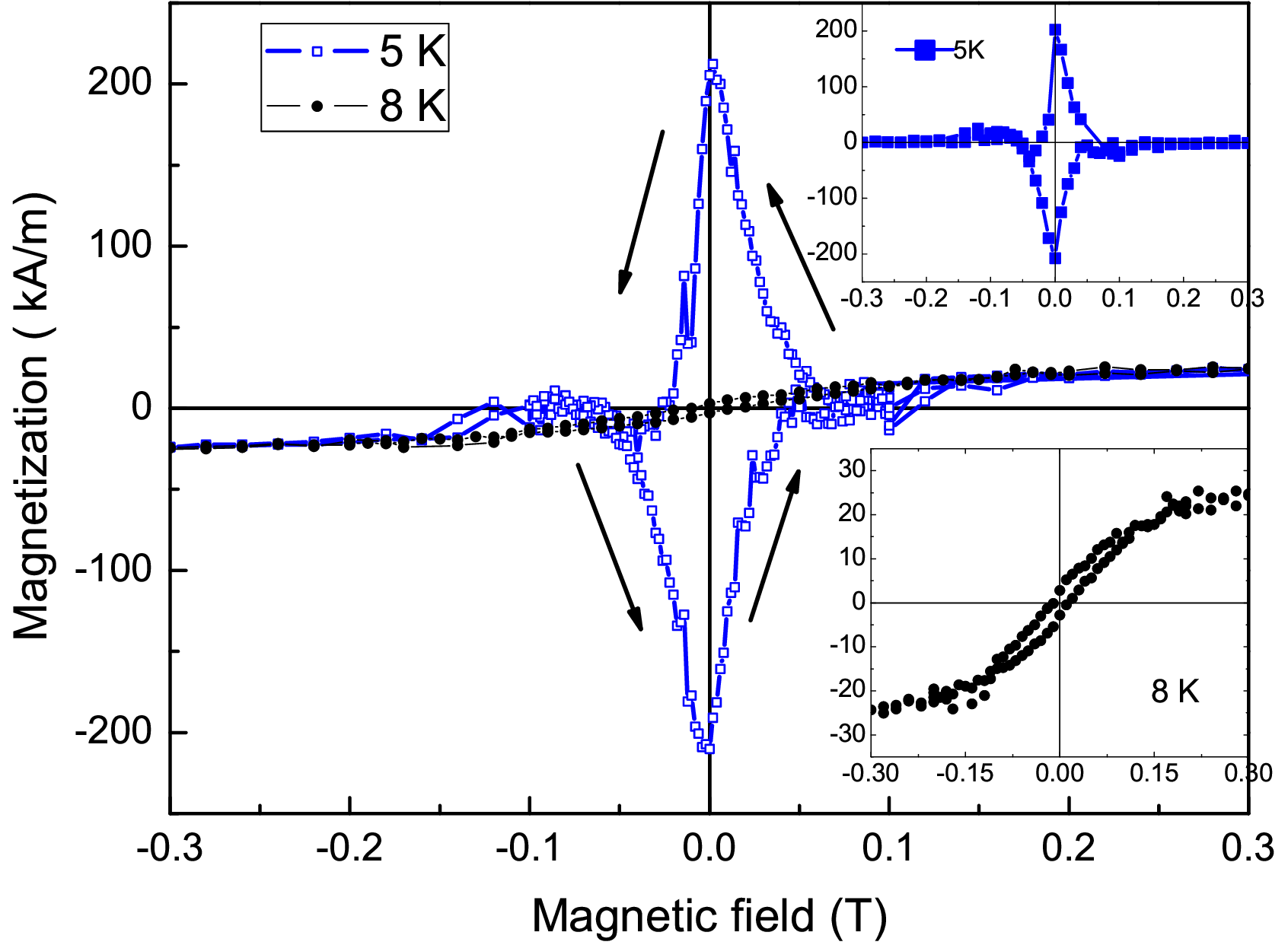}
\caption{Hysteresis loops of Pb/Co nanocomposites at 8 K and 5 K, which are above and below $T_c$, respectively. The upper inset shows the hysteresis loop at 5 K with the ferromagnetic signal subtracted, i.e., it is only the superconducting signal. The lower inset shows the center part of the curve at 8 K. \label{fig-mhf}}
\end{figure}

The blocking temperature of the Co particles is around 25 K. Figure \ref{fig-mhf} shows the hysteresis loops for the sample at 5 K and 8 K and we can see that the sample is ferromagnetic at 8 K as expected. The hysteresis loop at 5 K, however, clearly shows the co-existence of SC and FM below H$_{c2}$, which is 0.15 T for this sample. Above H$_{c2}$, the system is ferromagnetic. The comparison of the hysteresis loop with the two insets in Fig. \ref{fig-mhf} reveals that when the sample enters the SC state, the magnetic moments of the Co nanoparticles ($\mu_{Co}$) are shielded and the superconducting signal is much stronger than the original ferromagnetic signal.

\begin{figure}[h]
\includegraphics[width=0.6\columnwidth]{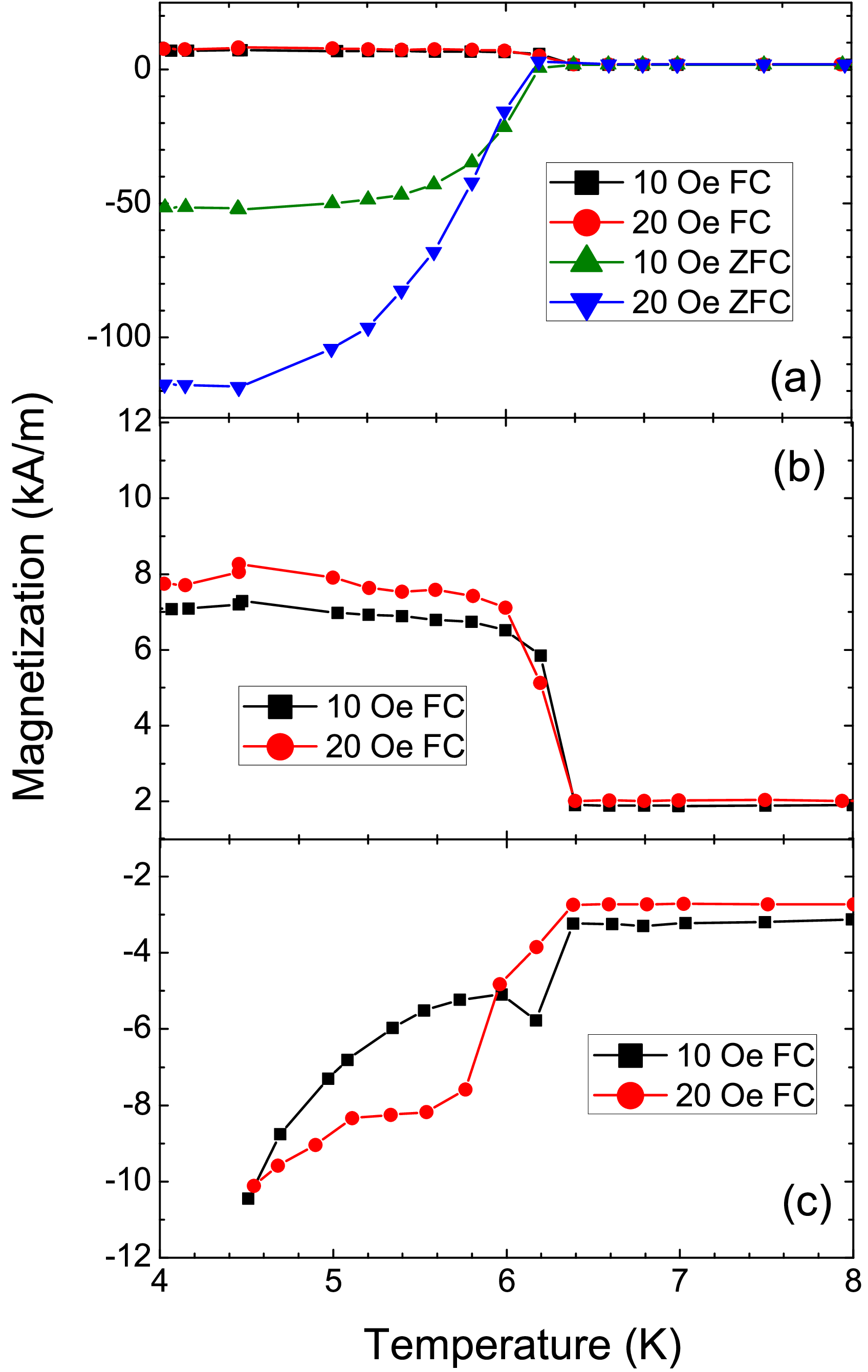}
\caption{DC magnetization as a function of temperature of the Pb/Co nanocomposites (a) ZFC and FC in positive external magnetic field from 300K to 4 K (b) enlargement of the FC curves and (c) FC in a negative magnetic field from 300K to 8K and then FC in positive external field to 4.5 K. \label{fig-mt}}
\end{figure}

Fig. \ref{fig-mt} shows the magnetization as a function of temperature (M-T curves) in a small external magnetic field. One can see in Fig. \ref{fig-mt} (a) from the ZFC M-T curves that the sample is in the diamagnetic Meissner state below T$_c$  with a transition to the normal state when the temperature goes above $T_c$. The decrease of T$_c$ from 7.2 (corresponding to pure Pb) to 6.2 K is due to the proximity effect and the spontaneous vortex formation in SC/FM hybrids and the details of this effect are discussed elsewhere\cite{YTXing08}. The FC M-T curves, shown in Fig. \ref{fig-mt} (a) and (b), have a positive signal which is known as PME in both conventional and high T$_c$ superconductors. When T goes above T$_c$, the PME signal disappears and the positive magnetization due to $\mu_{Co}$ is left.

From Fig. \ref{fig-mt} (a) and (b) one cannot conclude that the PME is due to the interaction between the SC Pb matrix and the embedded FM Co particles. For that reason, we performed further measurements using the following procedure: first we aligned $\mu_{Co}$ by applying a negative external magnetic field of 0.5 T above the blocking temperature ($T_b$) and cooled down the sample to 7 K which is below $T_b$ of the Co particles but above T$_c$ of the Pb matrix. Then the negative magnetic field is removed and a small positive field (10 Oe and 20 Oe, respectively), which is too small to switch the direction of the aligned magnetic moments, is applied in order to do the same FC measurements as shown in Fig. \ref{fig-mt} (b). The surprising result is shown in Fig. \ref{fig-mt} (c): the PME in the FC M-T curves [seen in Figs. \ref{fig-mt} (a) and (b)]  disappears and instead a DME is observed. The difference between the two measurements is only the orientation of $\mu_{Co}$ relative to the external field below T$_c$: when the direction of $\mu_{Co}$ is opposite (antiparallel) to the external field one has a DME [Fig. \ref{fig-mt} (c)] while one has a PME when direction of $\mu_{Co}$ is parallel to the field. Furthermore, we can conclude that the direction of the Co magnetic moments has not been switched by the vortices induced by positive external magnetic field in the SC, otherwise the signal in Fig. \ref{fig-mt} (c) would be positive rather than negative above T$_c$. This conclusion will be used in the discussions below.

In the following we will give an explanation for the observation of the PME in our Pb/Co nanocomposites. First we can rule out that the positive PME signal is just the sum of the negative DME signal of the SC Pb matrix and a stronger positive FM signal of the $\mu_{Co}$. The SQUID signal in Fig. 2 (b) decreases suddenly at T$_c$ by a factor of about 3 when the temperature goes above T$_c$ in the warm-up process. On one hand it means that the PME signal is strongly associated with the superconducting transition and, on the other hand, there is no reason why the positive contribution of the $\mu_{Co}$ should increase by such a large factor below T$_c$. As discussed before, the fact that changing the orientation of $\mu_{Co}$ relative to the direction of the external field produces a change between PME and DME, clearly indicates that the PME is due to the interaction between the SC Pb matrix and the FM Co particles.

In our previous work~\cite{YTXing08} we have shown that the sample is a type II SC with a coherence length $\xi$ being somewhat larger than the diameter d of the FM particles. The magnetic stray field of ferromagnetic particles inside a type II SC can lead to spontaneous vortices in different forms, such as vortex-antivortex pairs, loops and even closed loops as predicted by theoretical calculations using Ginzburg-Landau theory~\cite{MMDoria07,MMDoria04}. These calculations, however, consider the case $d \gg\xi$, and they do not apply to our sample where $d\le \xi$. 

Since our Co particles are single domain particles, the magnetic moment of each particle has to be suppressed and shielded by the SC matrix inducing a supercurrent, i.e. forming either just a non-superconducting sphere (\textquotedblleft point-vortex\textquotedblright) around the particle or the type of spontaneous vortices as described above. Because our samples contain a large number of magnetic particles, there will be many of these vortices forming a random network. A detailed transport study of such a vortex state in Pb/Co nanocomposites can be found in our previous work \cite{YTXing08}. Here we want to concentrate on the magnetic properties of this vortex state. We focus on the magnetization due to vortices generated by i) randomly oriented magnetic moments of the Co particles, ii) oriented Co moments and iii) the application of an external magnetic field.

In the ZFC process, going from room temperature below T$_c$, all the $\mu_{Co}$ are randomly oriented resulting in a zero magnetization of the Co magnetic moments, i.e. $M_{Co}$ = 0. Using the well-known relationship between the total magnetic field $B_{TOT}$, the applied external magnetic field H and the magnetization M
\begin{equation}
B_{TOT} = \mu_0 (H + M) \label{equ2}
\end{equation}
we obtain  $B_{TOT} = 0$ for the sample without external magnetic field above T$_c$. Applying a small magnetic field H below T$_c$, all the external field or most part of it is expelled out of the sample due to the Meissner effect, resulting in $B_{TOT} \approx 0$. Using equation (\ref{equ2}) we, therefore, can conclude that the sample has a negative response (magnetization M) to the external magnetic field H., i.e. shows the usual, well-known DME.

In the FC process, on the other hand, the situation is quite different. The external magnetic field will align $\mu_{Co}$ when T goes below $T_b$. In addition, cooling further down, the magnetic field will be trapped in the sample in form of vortices when T goes below T$_c$. We can divide the total magnetic field $B_{TOT}$, below T$_c$ in two parts: the external field induced part, $B_{EF}$,and the Co particle induced part, $B_{IF}$, where IF means induced flux:

\begin{equation}
B_{TOT} = B_{EF} + B_{IF} \label{equ3}
\end{equation}
Due to the Meissner effect, the external magnetic field $H$ usually is expelled from the sample if $H<H_{c1}$, $H_{c1}$ being the critical field for the pure Meissner state. In our case, however, due to the FM Co particles we have the formation of spontaneous vortices below $T_c$, i.e. we essentially have $H_{c1}\sim 0$. Furthermore, the Co particles work as pinning centers and trap the external magnetic field in most of the sample ($H$ only is expelled near the surface). Consequently, $B_{EF}$ is very close to $\mu_0 H$. For the FC process, the magnetic moments $\mu_{Co}$ are aligned and $B_{IF} > 0$. Consequently we have $B_{TOT} = B_{EF} + B_{IF} > \mu_0 H$ and using equation (\ref{equ2}) we obtain $M = B_{TOT}/\mu_0 - H > 0$, i.e. a paramagnetic signal. According to this analysis, the paramagnetic signal below $T_c$, i.e. the observed PME in the FC measurements, results from the Co-induced vortices. From the above discussion, we can easly understand why we produce a change from PME to DME by aligning $\mu_{Co}$ in opposite direction to the applied positive external field as shown in Fig. \ref{fig-mt} (c). If the magnetic moments are aligned in the direction opposite to the field, $B_{IF}<0$ and $B_{TOT} = B_{EF} + B_{IF} < \mu_0 H$. Again, using equation (\ref{equ2}) we obtain $M<0$ that gives rise to a diamagnetic response. It is important to point out that the origin of this DME is different from the typical DME response. In a normal SC, the origin of the DME is the Meissner effect, while here, it is mainly induced flux coming from the Co particles.  The magnitude of the DME signal in Fig. \ref{fig-mt} (c) is somewhat larger than that of the PME signal in Fig. \ref{fig-mt} (b), indicating that not all the external field H is trapped, i.e. $(B_{EF}/\mu_0 - H)<0$ below T$_c$.

We now return to the ZFC measurements and ask the following question: Is it possible to observe the PME in ZFC experiments, i.e. without having a trapped external flux? In this case we have $B_{EF} = 0$ which gives $B_{TOT} = B_{IF}$ (see Eq. (\ref{equ3})) and the magnetization of the sample becomes
\begin{equation}
M = B_{TOT}/\mu_0 - H = B_{IF}/\mu_0 -H\label{equ4}
\end{equation}
From this equation, it follows that if $B_{IF} > \mu_0 H$ the sample will show PME even in a ZFC process and, vice versa, if $B_{IF} < \mu_0 H$ the sample will show a DME.

\begin{figure}[h]
\includegraphics[width=0.5\columnwidth]{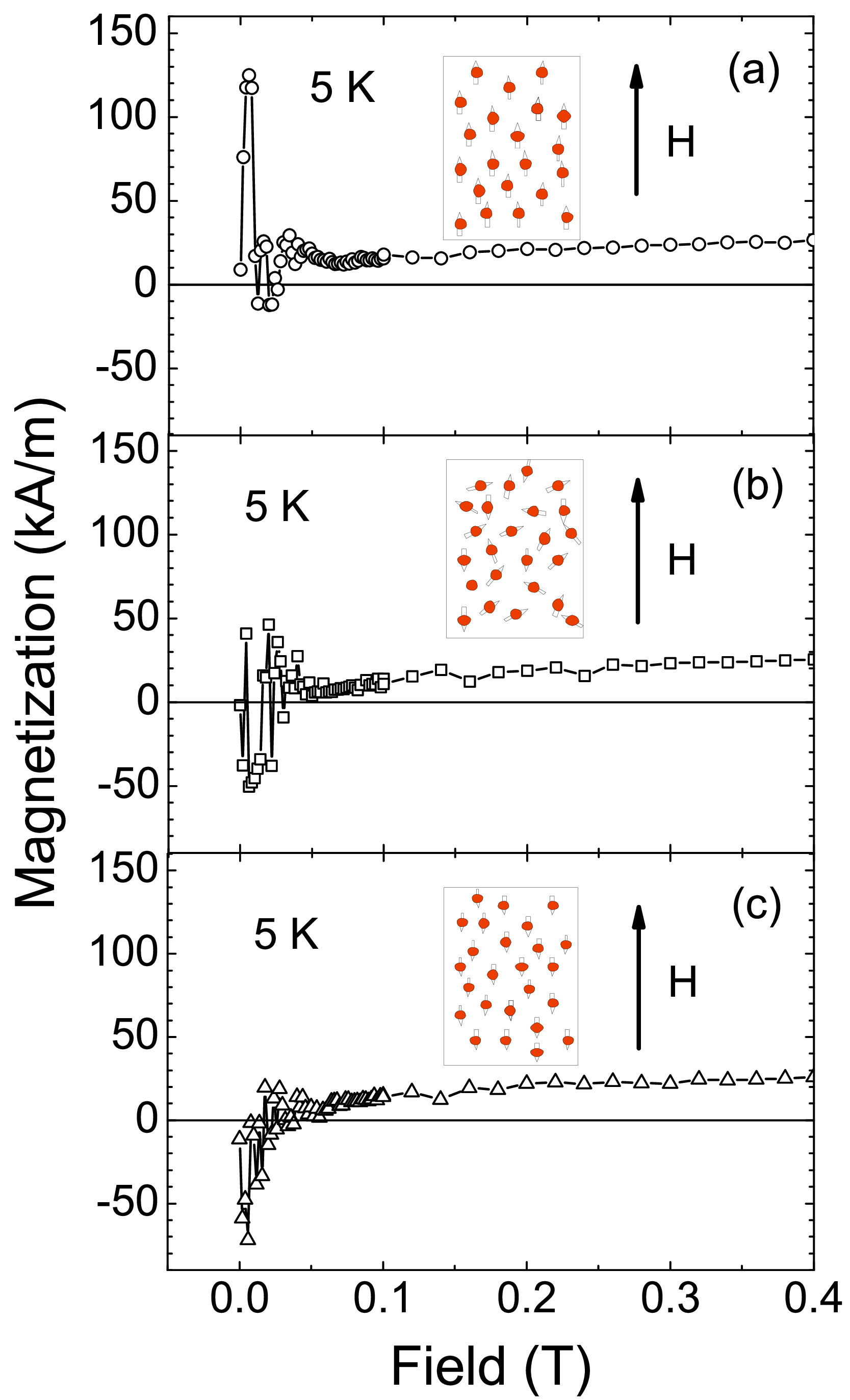}
\caption{The virgin magnetization curves of Pb/Co nanocomposites at 5 K with the moments of the Co particles (a) aligned to the direction of the external field (b) randomly oriented  and (c) aligned to the opposite direction of the external field. The insets show the direction of the internal moments in the sample during the measurements.} \label{fig-mhv}
\end{figure}

In order to test the above conclusions, the virgin magnetization curves have been measured for different orientations of $\mu_{Co}$ relative to the external magnetic field. Fig. \ref{fig-mhv} shows that the magnetization presents fluctuations as a function of the magnetic field only inside the superconducting region (H $<$ H$_{c2}$). This fluctuations are an indication of vortex rearrangement~\cite{YTXing08} and will not be discussed in details here. Although the data below H$_{c2}$ are too noisy to give a quantitative analysis, it can give a qualitative estimation of the magnetic property. The difference in the three plots of Fig. \ref{fig-mhv} is quite clear: from Fig. \ref{fig-mhv} (a) one can find that in the low field region the sample indeed has a positive signal when applying a field parallel to the direction of the aligned Co particles, i.e. parallel to $\mu_{Co}$. If all the moments are randomly oriented the total moment of all Co particles is zero. It results in $B_{IF} = 0$ and the magnetization of the sample becomes $M \approx -H$. We can see in Fig.\ref{fig-mhv} (b) that most of the data points for small magnetic fields indeed are negative. If $\mu_{Co}$ are aligned antiparallel to the external field, $B_{IF} < 0$. According to equation (\ref{equ4}) the sample should have a more negative moment than for the randomly oriented sample, which is confirmed by the data shown Fig. 3(c). These results clearly show that in the Pb/Co nanocomposites one can manipulate PME and DME even in ZFC experiments which to our knowledge has not been reported before.

We want to emphasize that the absolute value of the magnetization at low fields is much larger than the value of the saturation magnetization resulting from the aligned Co magnetic moments. This indicates that the spontaneous vortices are the origin of the magnetization at low fields. Furthermore, it shows that the transition into the superconducting state gives an amplification of the magnetization due to the formation of spontaneous vortices having a much larger magnetic moment than the Co particles. In other words, the interaction between the SC matrix and the aligned FM particles embedded in the matrix leads to a surprising amplification of the sample magnetic moment below T$_c$.

There are few points we should comment about our results. First, all the discussion above is only valid when the external field, applied after alignment of $\mu_{Co}$ with a large field of opposite direction, is small enough not to re-align $\mu_{Co}$ . In this case it does not change the original orientation of the spontaneous vortices in the sample. Second, when the magnetic field becomes larger, the interaction between the external field and the spontaneous vortices becomes more complicated since the external field will penetrate the SC in the form of vortices that can change the stray field of the Co particles and consequently, change the form of the spontaneous vortices.

In conclusion, our studies of the magnetic properties of Pb/Co nanocomposites show that these samples have unusual properties in an external magnetic field. For the first time we achieved the manipulation of PME and DME in both FC and ZFC measurements by changing the orientation of the magnetic moments of the Co nanoparticles relative to the external magnetic field. We find clear evidence for the formation of spontaneous vortices, induced by the Co particles, being the reason for the PME observed in this work. These vortices are intrinsically different from the ones created by magnets lying outside the SC~\cite{JIMartin97,MLange05,ZRYang04}. Due to these spontaneous vortices, the SC/FM nanocomposite has novel magnetic behavior which has not been observed before. These findings, therefore, give an important contribution to the study of the interplay between SC and FM besides the already observed ones, such as proximity effect~\cite{AIBuzdin05}, domain wall superconductivity~\cite{ZRYang04} and hysteresis pinning~\cite{APalau07}, etc.

This work was partially supported by CAPES/DAAD cooperation program and the Brazilian agencies CNPq, FAPERJ (Cientistas do Nosso Estado and  PRONEX) and L'Oreal Brazil. H. Micklitz acknowledges CAPES/DAAD and PCI/CBPF for financial support.

%
%

\end{document}